\documentclass[11pt]{article}
\usepackage{graphicx}
\usepackage{subfigure}
\usepackage{mathrsfs}

\usepackage[english]{babel} 

\usepackage{blois}
\DeclareGraphicsExtensions{.pdf,.png,.jpg,.eps} 
\graphicspath{{figs/}}

\def\Journal#1#2#3#4{{#1} {\bf #2}, #3 (#4)}

\def\PLB{{\em Phys. Lett.}  B}
\def\PRL{\em Phys. Rev. Lett.}
\def\PRD{{\em Phys. Rev.} D}

\def\JHEP{{\em JHEP} }

\def\be{\begin{equation}}
\def\ee{\end{equation}}
\def\bea{\begin{eqnarray}}
\def\eea{\end{eqnarray}}

\newcommand{\s}{\ensuremath{\sqrt{s}}}

\newcommand{\pp}{\ensuremath{pp}}

\newcommand{\pt}{\ensuremath{p_{\rm T}}}

\newcommand{\aS}{\ensuremath{\alpha_{s}}}

\newcommand{\Photo}{}

\author{L. BRYNGEMARK, FOR THE ATLAS AND CMS COLLABORATIONS}
\address{Lund University, Department of Physics, Particle Physics, \\Box 118, S-221 00. Lund, Sweden}
\title{QCD JET RESULTS IN ATLAS AND CMS}

\begin{document}
%\linenumbers
\vspace*{4cm}

\maketitle

\abstracts{A brief summary of recent measurements of QCD jet production at the LHC, featuring results from ATLAS and CMS, is given. Results for heavy flavours are also shown.}

\section{Introduction}
Jets are abundant in the final states produced in proton-proton (\pp) collisions at the Large Hadron Collider (LHC). They probe many aspects of Quantum Chromodynamics (QCD), such as the proton structure (parton distribution function, PDF) and the strength of the strong coupling constant \aS{}, and provide an environment for testing perturbative QCD (pQCD). 

Some of the recent QCD measurements made with jets in the final state by ATLAS\cite{detectorpaper} and CMS\cite{cmsDetectorpaper} are presented here, focusing on the physics conclusions. Details on the analyses can be found in the references. The detectors are described elsewhere.\cite{detectorpaper,cmsDetectorpaper}
 For jet measurements, apart from the calorimeter, tracking and muon systems are also used. Energy calibration gives the full energy in a jet, considering energy lost in hadronic interactions, upstream of the calorimeter, due to thresholds etc. 
A major experimental challenge for jet measurements is pile-up (overlaid signal from additional \pp{} interactions). Pile-up introduces fluctuations, distorting the energy measurement.
Tracking information can be used to remove charged particle contributions. ATLAS and CMS both use a method first proposed by Cacciari and Salam\cite{area}, in which the pile-up density is measured in the event and the jet area is used as an estimate of the pile-up susceptibility of each jet. Multiplied, these quantities give the amount of pile-up to subtract from the measured jet transverse momentum (\pt). This is the first step in the calibration chains.

\section{Jet measurements}
\subsection{Inclusive jet cross sections and PDFs}
\begin{figure}[!t]
\centering
\subfigure[]{\label{fig:ptSpec}\includegraphics[height=0.35\linewidth]{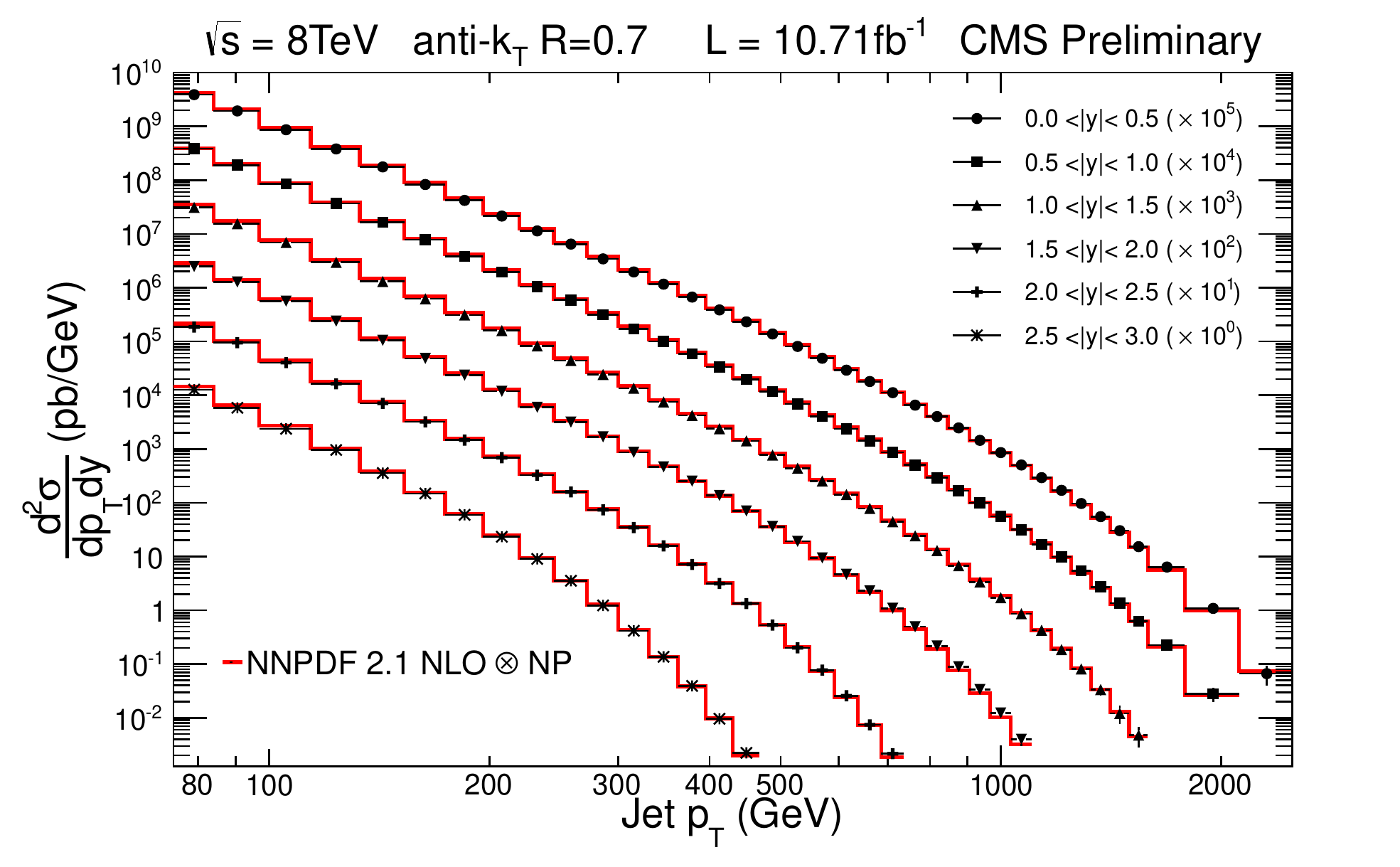}}
\subfigure[]{\label{fig:xG}\includegraphics[height=0.36\linewidth]{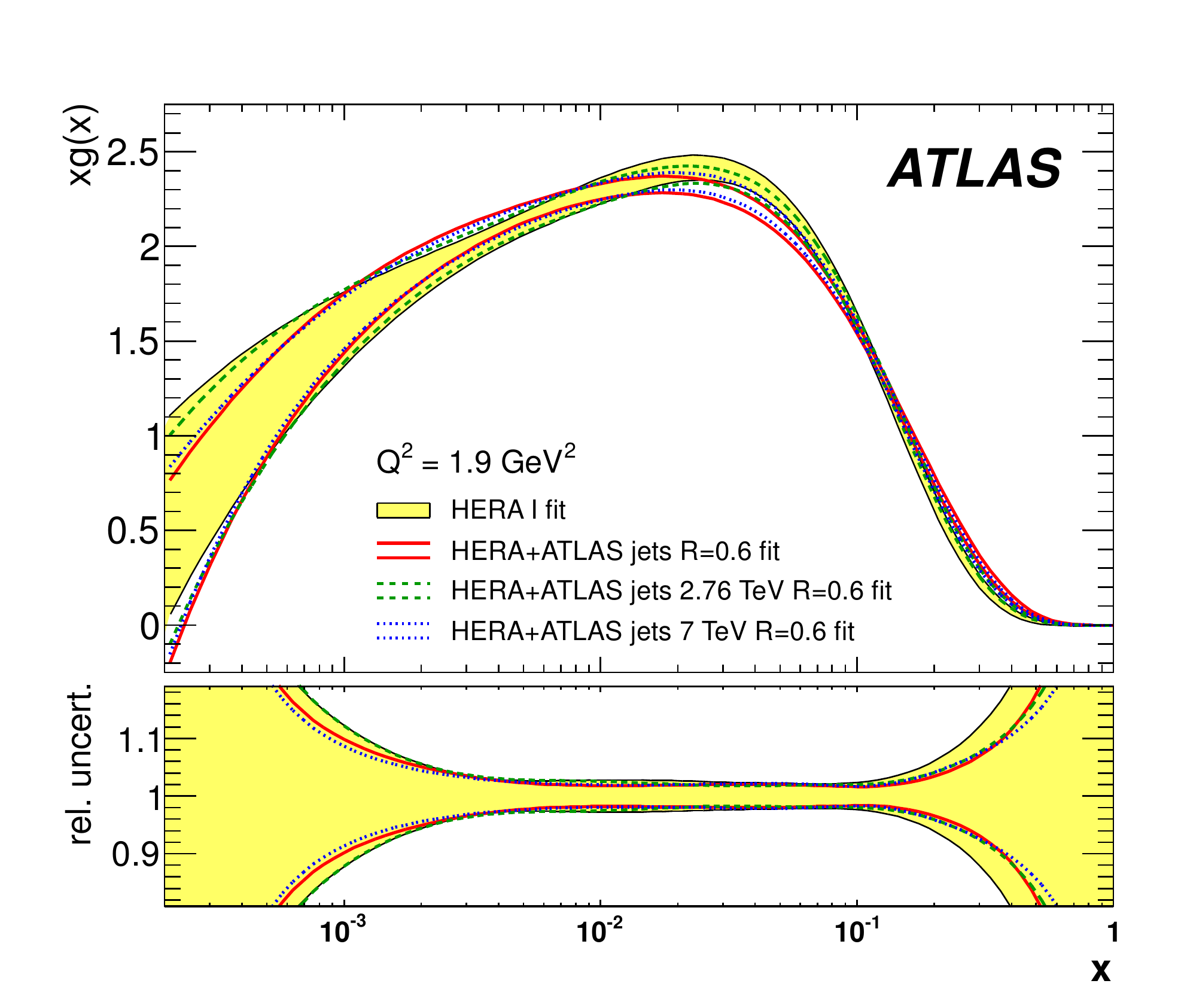}}\\
\caption{ Results for measurements of inclusive jet cross section.  \subref{fig:ptSpec} CMS: jet \pt{} and NLO$\otimes$NP prediction (line) in bins of rapidity $y$.\protect\cite{cmsInclJets8} \subref{fig:xG} Gluon $x$ distribution extracted using HERA data only (yellow band); combined with ATLAS jet data (red lines); combined with ATLAS 2.76 and 7~TeV data respectively (dashed, green and blue).\protect\cite{atlas2-76}}
\label{fig:inclJets}
\end{figure}

The inclusive jet cross section is calculable in pQCD, and can be used to constrain \aS{} and PDFs. It has been measured by CMS at centre-of-mass energies \s{} = 7 and 8~TeV.\cite{cmsInclJets7,cmsInclJets8} 
Comparison to next-to-leading order (NLO) Monte Carlo (MC) simulation with a non-perturbative correction factor (NP) shows good agreement over several orders of magnitude, as seen in Fig. \ref{fig:ptSpec}. 

ATLAS has performed a similar measurement\cite{atlas2-76} at both \s~=~2.76 and 7~TeV. With correlations between the two measurements taken into account in a data/NLO + NP double ratio, the experimental uncertainties are much reduced, and the data show preference for different PDF sets. Furthermore, the correlations have been exploited in a fit of gluon, valence and sea quark distributions to the data combined with deep inelastic scattering data from HERA\cite{hera}, as exemplified in Fig.~\ref{fig:xG}. The PDFs have been obtained using HERAfitter.\cite{hera,herafit,herafit2} The resulting gluon distribution is harder than that obtained from HERA data only, with reduced uncertainties at Bjorken $x > 0.1$.

\subsection{\aS}
\begin{figure}[!h]
\centering
\subfigure[]{\label{fig:aSfits}\includegraphics[height=0.37\linewidth]{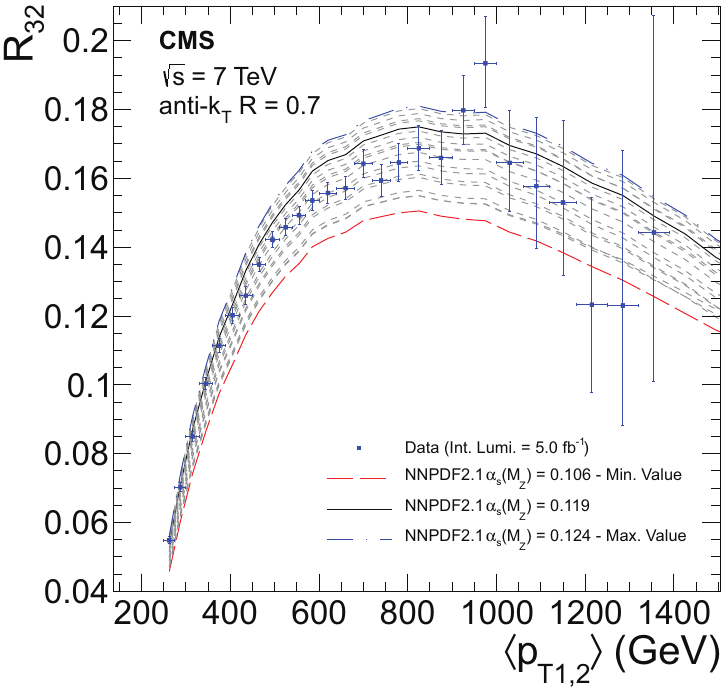}}
\subfigure[]{\label{fig:aSrun}\includegraphics[height=0.42\linewidth]{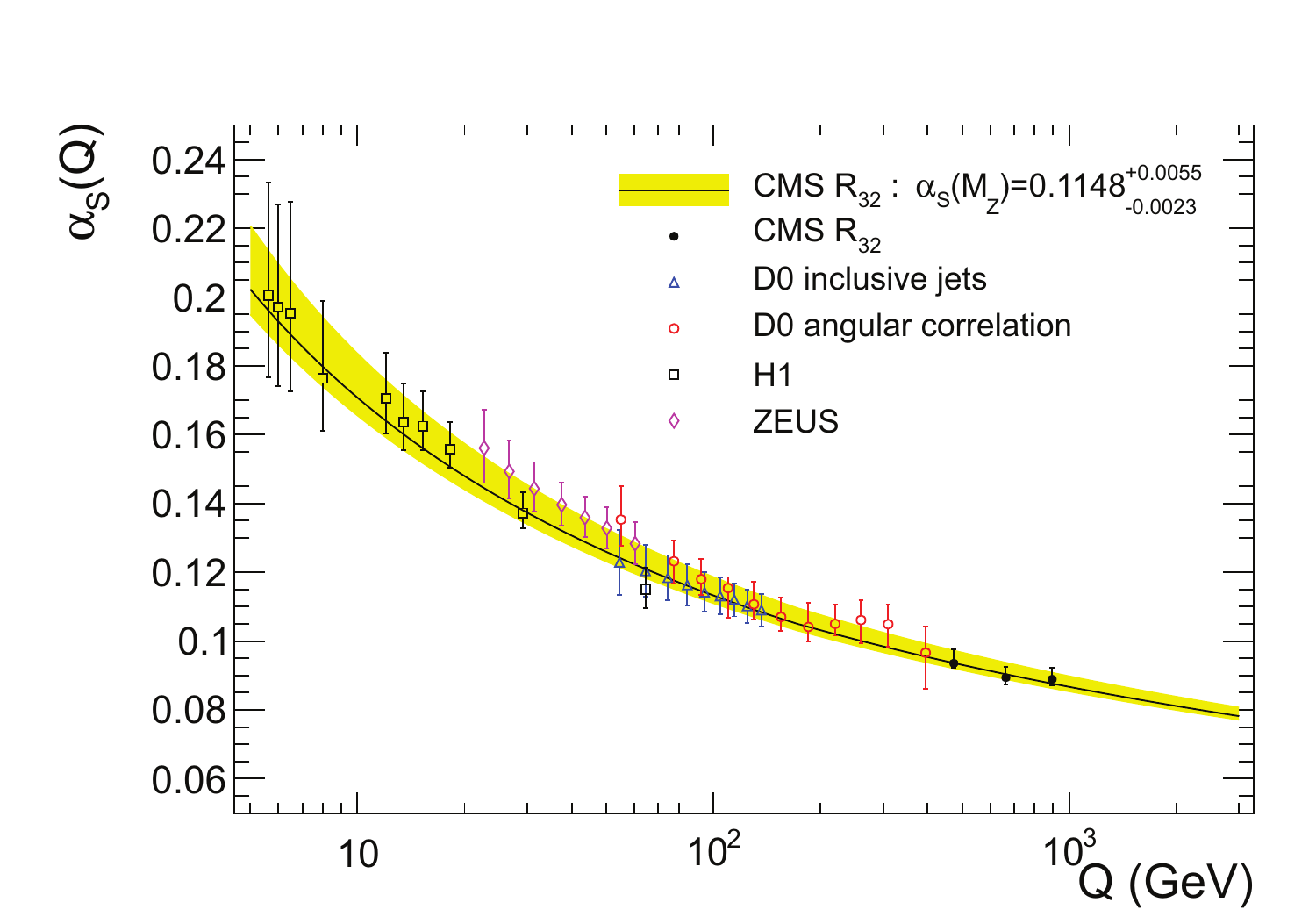}}\\
\caption{Measurements of \aS. \subref{fig:aSfits} \aS($M_Z$) of the NNPDF2.1 fit with the minimum $\chi^2$ is used for each PDF~set. \subref{fig:aSrun} Data from several experiments overlaid with the RGE prediction of \aS($Q$) from \aS($M_Z$) as measured by~CMS.\protect\cite{aScms}}
\label{fig:aS}
\end{figure}

The value of \aS{} is measured experimentally. Its strength is predicted to ``run'', i.e. depend on the energy scale at which it is probed. The evolution from a given starting point can be calculated using the Renormalisation Group Equation (RGE). At the LHC, a new energy regime is available for experimentally testing this evolution. Exploiting that the probability of radiating additional partons is proportional to additional powers of \aS, the cross section ratio 
\begin{equation}
R_{32} = \frac{\sigma \left (N_{jets} \geq 3\right )}{\sigma \left (N_{jets} \geq 2\right )}
\end{equation}
gives an observable that is proportional to \aS. $R_{32}$ is given in bins of $Q = \langle \pt \rangle =~\frac{ \pt^1+\pt^2 }{2}$, where $\pt^{1(2)}$ is the \pt{} of the (second) most energetic, (sub)leading, jet. A measurement by ATLAS of this ratio using the 39~pb$^{-1}$ 2010 data set,\cite{aSatlas} and by CMS using 5.0~fb$^{-1}$ from 2011,\cite{aScms} give \aS{} both at the reference scale $M_Z$ and for a range of $Q$. \aS~is extracted by fitting data with NLO PDFs, varying \aS$(M_Z)$ and choosing the fit with the minimum $\chi^2$. Figure \ref{fig:aSfits} shows fits to CMS $R_{32}$ data using NNPDF2.1, for \aS$(M_Z)$ between 0.106 and 0.124. 

ATLAS measures $\aS(M_Z)=0.111 \pm 0.006{\rm(exp)} _{-0.003}^{+0.016}{\rm(theory)}$. CMS measures $\aS(M_Z)=0.1148 \pm 0.0014{\rm(exp)}\pm 0.0018{\rm(PDF)}_{-0.0000}^{+0.0050}{\rm(scale)}$. In both cases, scale uncertainties dominate. Figure \ref{fig:aSrun} shows \aS($Q$)~in CMS data, overlaid with data from other experiments. The RGE prediction of the evolution from the CMS measurement of $\aS(M_Z)$ is in agreement with all data.

\subsection{Heavy flavour jets}
The production cross section of heavy flavour quarks ($b$ and $c$) should be approximately independent of low-scale hadronisation effects, owing to the high quark masses that are above the typical QCD scale $\Lambda_{\rm QCD}$. Hence production features of heavy flavour jets can probe QCD.

To first order, heavy quarks are pair produced in the hard interaction and should be of equal and opposite momenta. Measuring angular correlations of heavy quark jet pairs thus probes higher orders of pQCD. This has been explored in a measurement of $b$-jet pair angular correlations by CMS.\cite{cmsBjets} There are also non-perturbative production modes such as \emph{gluon splitting}, when an outgoing gluon splits into a heavy quark pair, and a non-pertrbative component to \emph{heavy flavour excitation}, when a sea quark participates in the hard interaction.

\begin{figure}[!h]
\begin{center}
\subfigure[]{\label{fig:bJetProd}\includegraphics[height=0.325\linewidth]{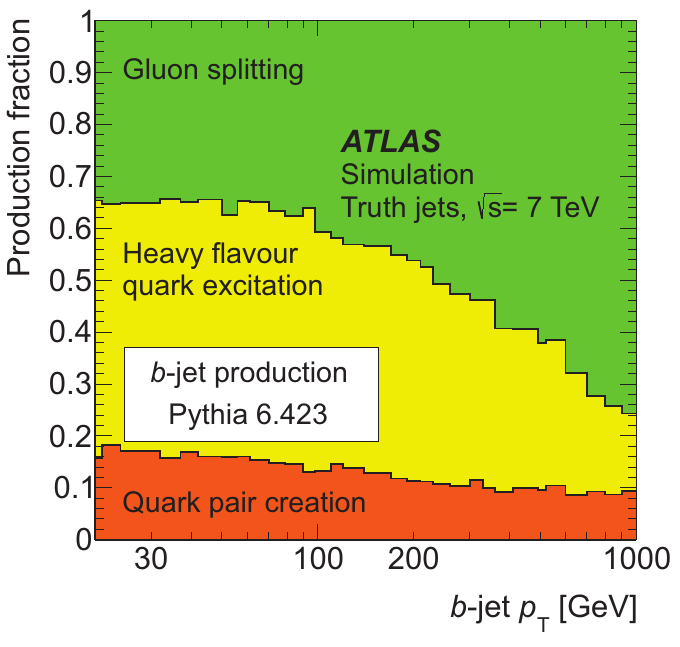}}
\subfigure[]{\label{fig:bJetAs}\includegraphics[height=0.33\linewidth]{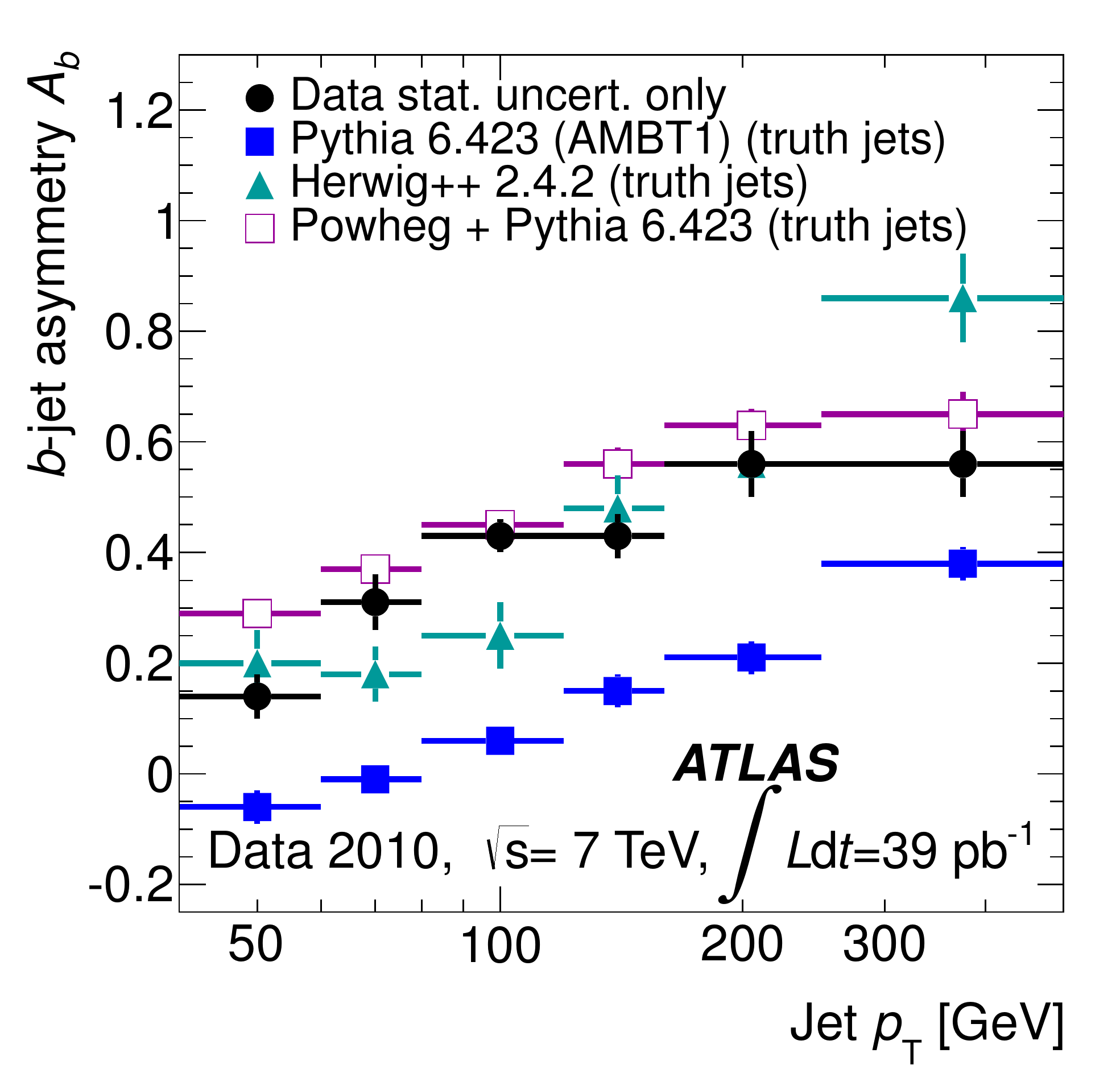}}
\subfigure[]{\label{fig:BUfrac}\includegraphics[height=0.33\linewidth]{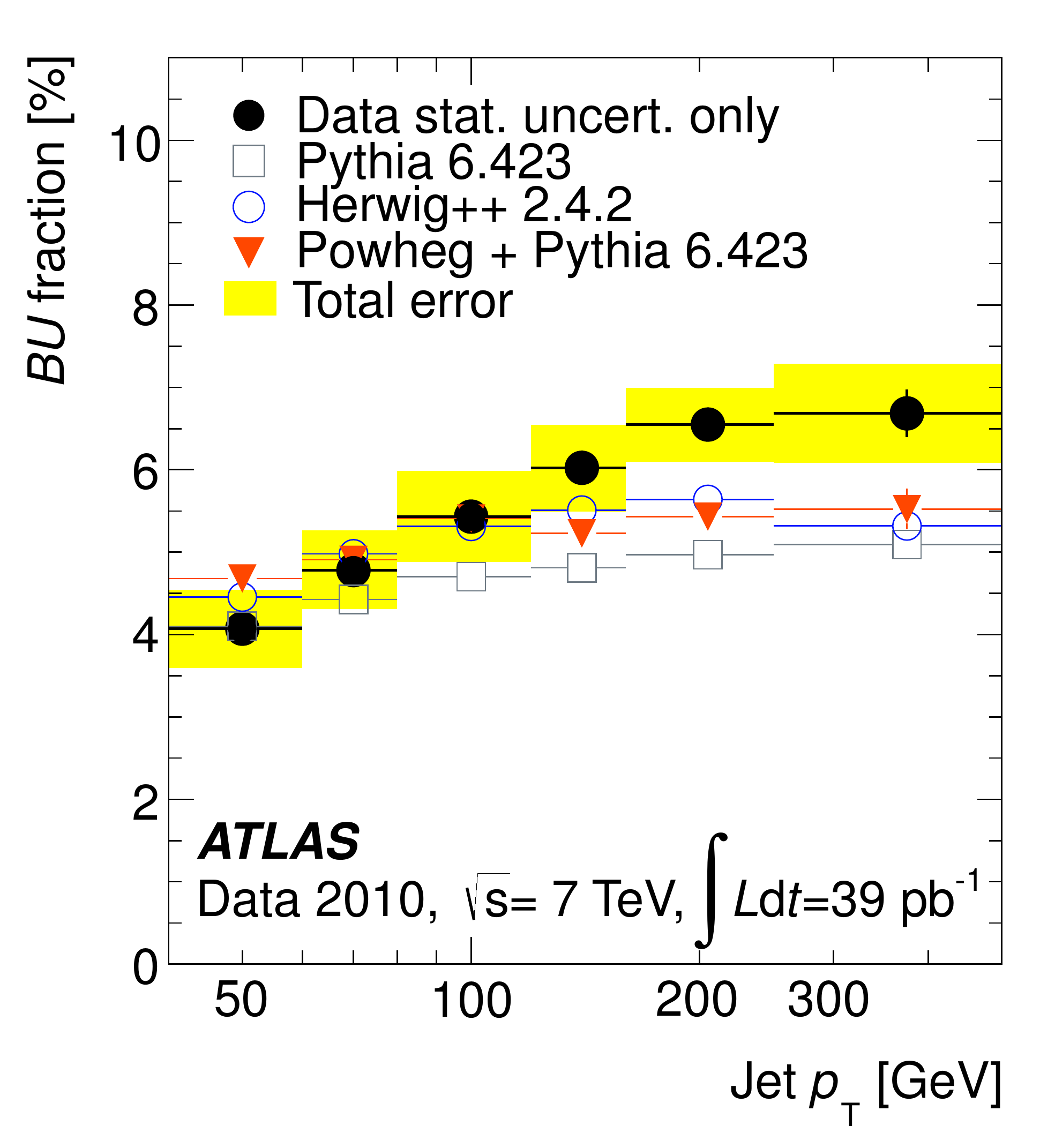}}\\
\caption{\subref{fig:bJetProd} {\sc Pythia} prediction of inclusive $b$-jet production modes. \subref{fig:bJetAs} Predictions of $b$-jet asymmetry vs $b$-jet \pt{} compared to data. \subref{fig:BUfrac} Fraction of jet pairs consisting of a light jet (``\emph{U}'') and a $b$-jet,  vs leading jet \pt.\protect\cite{atlasBjets}}
\label{fig:bjetATLAS}
\end{center}
\end{figure}

ATLAS has measured the dijet flavour fractions\cite{atlasBjets} in the 39~pb$^{-1}$ 2010 data set. The different kinematics of secondary vertices in heavy flavour decays compared to those of particles containing strangeness is exploited to identify $b$- and $c$-jets, using a template fit. The leading jet pairs are categorised according to their composition in terms of light ($u, d, s$ or $g$), $c$- or $b$-jets. Allowing mixed-flavour pairs gives access to all production modes, whose proportions are seen in Fig. \ref{fig:bJetProd}. Due to energy lost in decays to neutrinos, or misreconstruction of jets from gluon splitting\footnote{merging or missing one of the produced $q\bar{q}$ jets gives a gluon splitting contribution to mixed flavour dijets}, $b$-jets tend to be subleading in mixed-flavour pairs. The $b$-jet asymmetry\footnote{The corresponding quantity is also defined for $c$-jets.},
\begin{equation}
A_{b} = \frac{N_{subleading}^{b}}{N_{leading}^{b}} - 1,
\end{equation}

\noindent where $N_{(sub)leading}^{b}$ is the number of (sub)leading $b$-jets, is shown versus $b$-jet \pt\ in Fig. \ref{fig:bJetAs}. In particular the LO generator {\sc Pythia} underestimates $A_{b}$.
The fraction of $b$ + light dijets is shown in Fig. \ref{fig:BUfrac}. This final state probes the non-perturbative effects of heavy flavour excitation or gluon splitting; it is also the only state where MC does not agree with data within uncertainties.

\subsection{Jet substructure}
Partons produced in a heavy particle decay receive a Lorentz boost if their energy is much larger than their mass. The resulting jets become increasingly collimated with larger boost, and may not be resolved as separate jets. This introduces \emph{boosted jets} with substructure, which can be exploited for event selection and reconstruction of these heavy particles. It becomes highly relevant to understand the structure of QCD jets, which are the background in analyses using boosted jets. In, for example, resonance searches, it is essential to understand how well MC models describe the QCD jet mass. 

\begin{figure}[!ht]
\begin{center}
\subfigure[]{\label{fig:dijUngroomed}\includegraphics[height=0.52\linewidth]{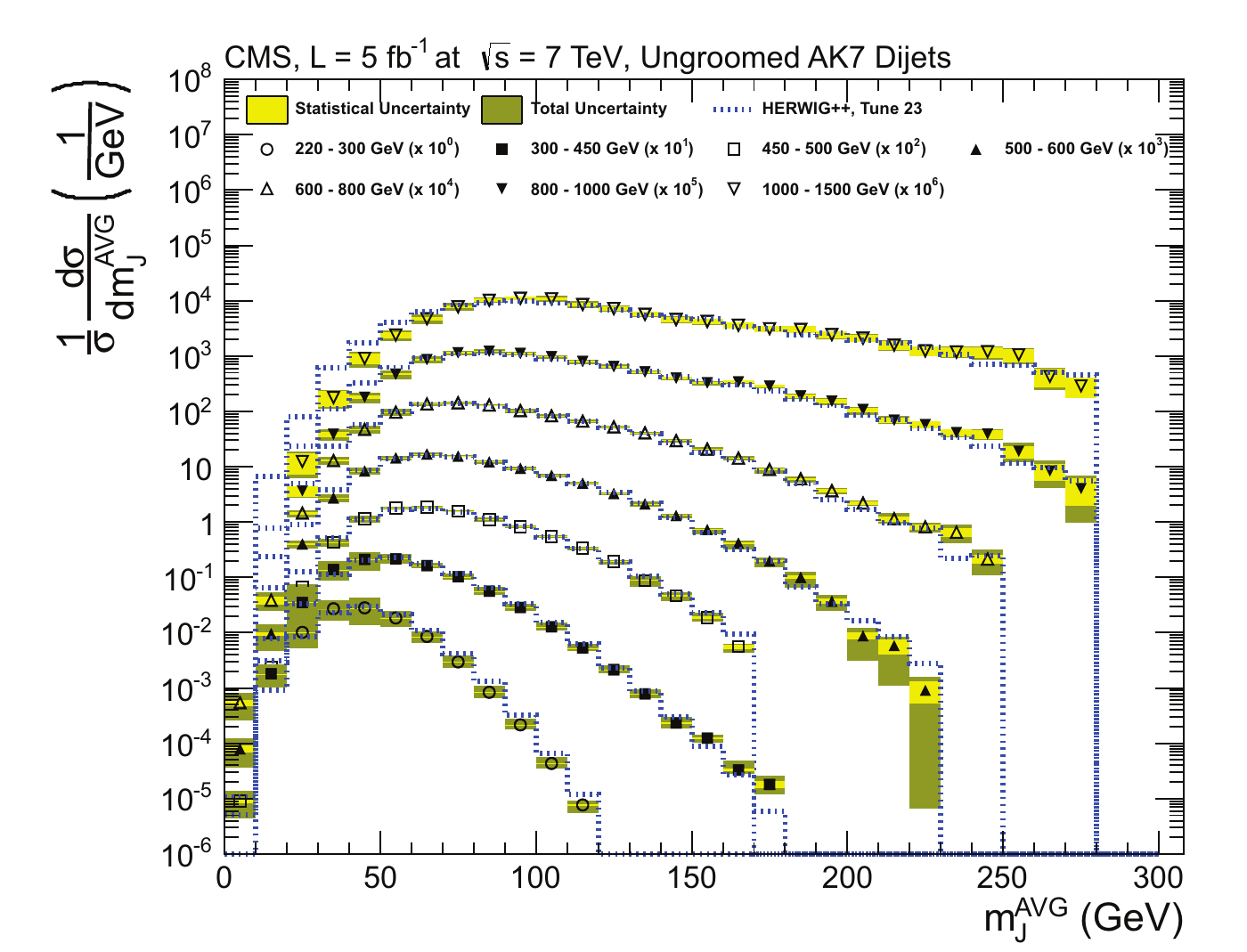}}
\subfigure[]{\label{fig:dijTrimmed}\includegraphics[height=0.52\linewidth]{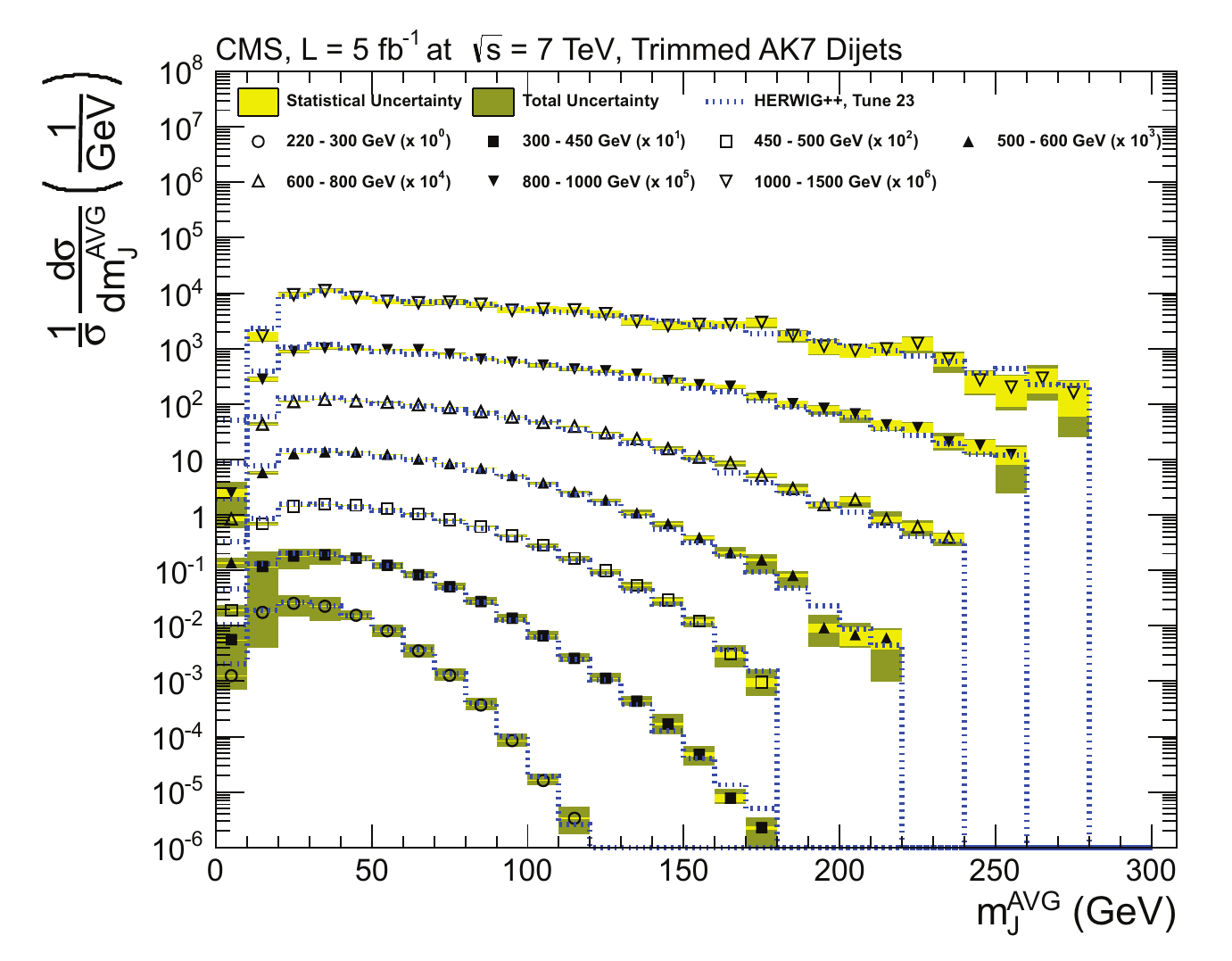}}\\
\caption{Leading jet pair average mass spectra in bins of \pt, \subref{fig:dijUngroomed} before and \subref{fig:dijTrimmed} after trimming. Data are overlaid with the prediction from Herwig++ (blue dashed line).\protect\cite{cmsSubs}}
\label{fig:dijetMass}
\end{center}
\end{figure}

Several substructure methods exist\cite{trim,filt,prun,prun2} to distinguish boosted jets from QCD brems-strahlung jets. CMS and ATLAS have both made several measurements\cite{atlasSubs1,atlasSubs2,cmsSubs} on jet substructure variables and the performance of different algorithms. One example is the impact of \emph{trimming}\cite{trim} on the average mass of the leading jet pair, shown in Fig. \ref{fig:dijetMass}, from a CMS analysis using the $5.0 \pm 0.2$~fb$^{-1}$ 2011 data.\cite{cmsSubs}
Comparing Figs. \ref{fig:dijUngroomed} and \subref{fig:dijTrimmed}, it is evident that trimming shifts the mass spectrum to lower values. Both this and the improved agreement between data and MC (Herwig++) at low masses is due to removal of soft contributions from pile-up and the underlying event at large angles; these are to some extent non-perturbative effects which are difficult to model.
Thus trimming makes the QCD predictions more robust. Finally the good agreement in the high-mass region is not destroyed by trimming, which is important since physics phenomena beyond the Standard Model in many cases (e.g. contact interactions) are expected to be visible in the high-mass tail of the spectra.

\section{Conclusions}

Results from some of the jet measurements by ATLAS and CMS have been shown. They show that jet measurements are precision measurements which provide important input for understanding the strong interaction. Jet data included in PDF fits can impact the extracted gluon momentum distribution. Measurements of \aS{} in a new energy regime are compatible with the evolution predicted by the RGE. Some discrepancies with respect to in particular LO generators are seen in jet $b$-hadron content. QCD jet substructure is well described by theoretical predictions, paving the way for beyond Standard Model discoveries.


\section*{References}


\begin{thebibliography}{99}
 
\bibitem{detectorpaper}ATLAS Collaboration, \Journal{\em JINST}{3}{ S08003}{2008}

\bibitem{cmsDetectorpaper}CMS Collaboration, \Journal{\em JINST}{3}{ S08004}{2008}

\bibitem{area}M. Cacciari and G. P. Salam,  \Journal{\PLB}{659}{119--126}{2008}

\bibitem{cmsInclJets7}CMS Collaboration, \Journal{ \PRD}{87}{12002}{2013} 

\bibitem{cmsInclJets8}CMS Collaboration, CMS-PAS-SMP-12-012, http://cds.cern.ch/record/1547589

\bibitem{atlas2-76}ATLAS Collaboration, \Journal{ {\em Eur. Phys. J.} C}{73}{2509}{2013}, arXiv:1304.4739

\bibitem{hera}H1 and ZEUS Collaboration, F. D. Aaron et al., \Journal{\JHEP}{01}{109}{2010}, arXiv:0911.0884

\bibitem{herafit}HERAFitter, http://projects.hepforge.org/herafitter.

\bibitem{herafit2}H1 Collaboration, F. Aaron et al. \Journal{{\em Eur. Phys. J.} C}{64}{561–587}{2009}

\bibitem{aSatlas}ATLAS Collaboration, ATLAS-CONF-2013-041, http://cds.cern.ch/record/1543225

\bibitem{aScms}CMS Collaboration, arXiv:1304.7498, submitted to {\em Eur. Phys. J.} C

\bibitem{cmsBjets}CMS Collaboration, CMS-PAS-BH-10-019, http://cds.cern.ch/record/1542457

\bibitem{atlasBjets}ATLAS Collaboration, \Journal{ { \em Eur. Phys. J.} C}{73}{2301}{2013}


\bibitem{trim}D. Krohn, J. Thaler and L.-T. Wang, \Journal{\JHEP}{1002}{084}{2010}

\bibitem{filt}J. M. Butterworth et al., \Journal{ \PRL}{100}{242001}{2008}

\bibitem{prun}S. D. Ellis, C. K. Vermilion, and J. R. Walsh,  \Journal{\PRD}{80}{051501}{2009}, arXiv:0903.5081

\bibitem{prun2}S. D. Ellis, C. K. Vermilion and J. R. Walsh,  \Journal{\PRD}{81}{094023}{2010}, arXiv:0912.0033

\bibitem{atlasSubs1}ATLAS Collaboration, ATLAS-CONF-2012-065, http://cds.cern.ch/record/1459530

\bibitem{atlasSubs2}ATLAS Collaboration, \Journal{\JHEP}{1205}{128}{2012}, arXiv:1203.4606

\bibitem{cmsSubs}CMS Collaboration,  \Journal{\JHEP}{05}{090}{2013}, arXiv:1303.4811


\end{thebibliography}
\end{document}